\documentclass{article}
\usepackage{spconf,amsmath,graphicx}
\usepackage{multirow}
\usepackage{threeparttable}
\usepackage{setspace}
\usepackage{amssymb}
\usepackage{siunitx}
\usepackage{booktabs}
\usepackage{enumitem}
\title{Acoustic-to-Articulatory Inversion based on Speech Decomposition and Auxiliary Feature}
\name{Jianrong Wang$^{1}$ \qquad Jinyu Liu$^{2}$ \qquad Longxuan Zhao$^{2}$ \qquad Shanyu Wang$^{2}$ 
\qquad Ruiguo Yu$^{1}$ \qquad Li Liu$^{3\star}$\thanks{$^{\star}$Corresponding author.}}
\address{$^{1}$ College of Intelligence and Computing, Tianjin University, Tianjin, China\\
$^{2}$ Tianjin International Engineering Institute, Tianjin University, Tianjin, China\\
$^{3}$ Shenzhen Research Institute of Big Data, the Chinese University of Hong Kong, Shenzhen, China}

\begin{document}
%\ninept
%
\maketitle
\begin{abstract}
\vspace{-0.2cm}
Acoustic-to-articulatory inversion (AAI) is to obtain the movement of articulators from speech signals. Until now, achieving a speaker-independent AAI remains a challenge given the limited data. Besides, most current works only use audio speech as input, causing an inevitable performance bottleneck. To solve these problems, firstly, we pre-train a speech decomposition network to decompose audio speech into speaker embedding and content embedding as the new personalized speech features to adapt to the speaker-independent case. Secondly, to further improve the AAI, we propose a novel auxiliary feature network to estimate the lip auxiliary features from the above personalized speech features. Experimental results on three public datasets show that, compared with the state-of-the-art only using the audio speech feature, the proposed method reduces the average RMSE by 0.25 and increases the average correlation coefficient by 2.0\% in the speaker-dependent case. More importantly, the average RMSE decreases by 0.29 and the average correlation coefficient increases by 5.0\% in the speaker-independent case.
\end{abstract}
\vspace{-0.4cm}
\section{Introduction}
\label{sec:intro}
\vspace{-0.35cm}
The conversion from acoustic speech to articulatory movement is called the acoustic-to-articulatory inversion (AAI) \cite{C1}, which plays a significant role in many applications (\textit{e.g}., pronunciation guidance \cite{C2}, helping patients with vocal or hearing impairments \cite{C3} and speech recognition \cite{C4}).

Early in \cite{C7}, the acoustic speech was mapped to articulatory movement with the codebook. However, the results of their inversion were highly dependent on the quality of the codebook. Then, with the publishing of corpora containing parallel acoustic and articulatory data, data-driven inversion frameworks based on machine learning were proposed. And the Mel-scale frequency cepstral coefficients (MFCC) of the speech signals were first accepted as inputs and then mapped to articulatory movements. Later, other methods like hidden markov model \cite{C8}, mixture density network \cite{C9}, and deep belief network \cite{C10} were proposed. Recently, with the rise of deep neural networks (DNNs), the deep bidirectional long short-term memory (DBLSTM) was used by \cite{C11} in the AAI. Since then, most of AAI related works (\textit{e.g}., \cite{C13,C14,C15,C34}) used DBLSTM to deal with various applications but the inputs of the models were only audio speech features. As for the speaker-independent AAI, the vocal tract length normalization \cite{C14} was proposed to transform the acoustic spaces of different speakers to a target one. \cite{C15} proposed the idea of pre-train and fine-tune to improve the generalization performance on their own dataset. Especially, \cite{C30} used one-dimensional convolution of different sizes to extract the audio feature. It improved the performance in speaker-independent case by adding extra phoneme information, achieving the state-of-the-art (SOTA) result on the public Haskins Production Rate Comparison (HPRC) \cite{C36} dataset. 

Until now, there are two main challenges leading to performance bottlenecks in AAI. Firstly, most of existing works only used the audio speech feature to predict the articulatory movement without exploiting any additional features (\textit{i.e}., lip feature, speaker identity feature or the content feature). Secondly, some works devoted to improve the generalization performance for the speaker-independent AAI, but their methods either lost the personalized information \cite{C14}, needed large 
amounts of data \cite{C15} or required additional phoneme information \cite{C30}.
\begin{figure*}[h]
	\centerline{\includegraphics[width=0.90\linewidth]{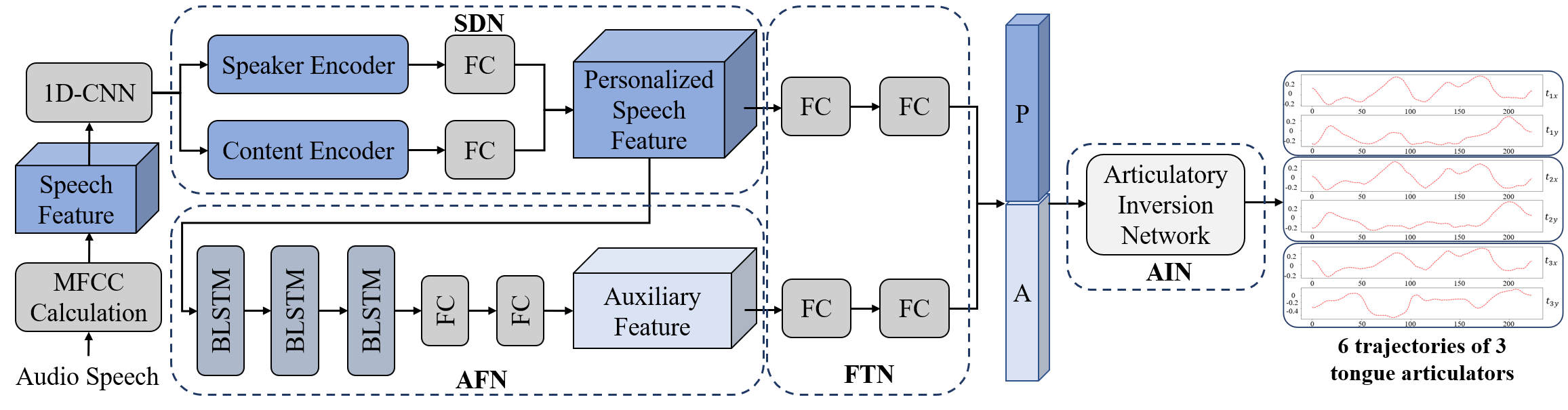}}
	\vspace{-0.3cm}
	\caption{The overview of our proposed SAFN. The P is the personalized speech features and the A is the auxiliary features.}
	\label{fig:ff1}
\end{figure*}

To address the above two challenges, we propose a novel SAF network composed of \textbf{S}peech Decomposition Network (SDN), \textbf{A}uxiliary Feature Network (AFN) and \textbf{F}eature Transformation Network (FTN), which we call SAFN in brief. Firstly, in order to adapt to the speaker-independent case, we explore a SDN inspired by the idea of speech synthesis \cite{C16} to obtain the personalized speech features. Then, to further improve the performance of AAI, we propose a novel AFN to estimate the lip auxiliary features as the prior information from the personalized speech features. Then, we design a FTN to generate feature pairs by transforming the personalized speech features and the lip auxiliary features. Last but not least, though AAI is not an one-to-one maping task (\textit{i.e}., different articulatory movements may correspond to the same speech), this problem can be alleviated by ensuring the smoothness of the articulatory movement on consecutive frames. An overview of our SAFN is shown in Fig. \ref{fig:ff1}.

In summary, this work has following three contributions: 1) To adapt to the speaker-independent case, a SDN is proposed to obtain the speaker embedding and the content embedding as personalized speech features. 2) To further improve the performance of AAI, a novel AFN is proposed to estimate the lip auxiliary features as prior knowledge. To the best of our knowledge, this is the first work that introduces an auxiliary feature instead of directly using the speech feature as input. 3) Speaker-independent AAI experimental results show the superior performance of the SAFN. On the public HPRC dataset, the SAFN outperforms the SOTA by a large margin (\textit{i.e}., the average RMSE decreases by 0.29 and the average  correlation coefficient increases by 5.0\%).
\vspace{-0.5cm}
\section{PROPOSED APPROACH}
\label{sec:method}
\vspace{-0.35cm}
\subsection{Overall Framework}
\vspace{-0.2cm}
In Fig. \ref{fig:ff1}, the core of SAFN are SDN and AFN. The SDN is pretrained to obtain the speaker embedding and the content embedding, which are served as the personalized acoustic features. The AFN is trained to estimate the corresponding lip auxiliary features as prior knowledge to help the prediction of tongue organ. To get such a network, the MFCC of the speech audio are extracted as speech features, and then the features are further encoded by multi-scale one-dimensional convolution. Besides, the speech features are sent to the pre-trained SDN to obtain the corresponding two embeddings, which are used as personalized speech features and sent to the AFN to obtain the lip auxiliary features. Later, the personalized speech features and the lip auxiliary features are feature fused \cite{C19} as multi-features. These multi-features are sent to the articulatory inversion network (AIN) to predict the movements of the tongue organs. The objective function for training is to minimize the combination of the reconstruction L2 loss of AFN and reconstruction L2 loss of AIN, which is given as:
\begin{equation}
\setlength{\abovedisplayskip}{3pt}
\setlength{\belowdisplayskip}{3pt}
	L = \alpha\times\sum_{i=0}^{m}\left(y_{l}^{i}-\hat{y}_{l}^{i} \right)^{2} + \beta\times\sum_{i=0}^{m}\left(y_{t}^{i}-\hat{y}_{t}^{i} \right)^{2},
\end{equation} 
where $\alpha$ and $\beta$ are set as 0.5 and 0.5 experimentally. $y_{l}^{i}$ and $y_{t}^{i}$ refer to the estimated lip auxiliary features by AFN and tongue movements by AIN, respectively. $\hat{y}_{l}^{i}$ and $\hat{y}_{t}^{i}$ refer to the corresponding real labels.

\vspace{-0.3cm}
\subsection{Speech Decomposition Network}
\begin{figure}[htb]
\begin{minipage}[b]{1.0\linewidth}
  \centering
  \centerline{\includegraphics[width=8.5cm,height=4cm]{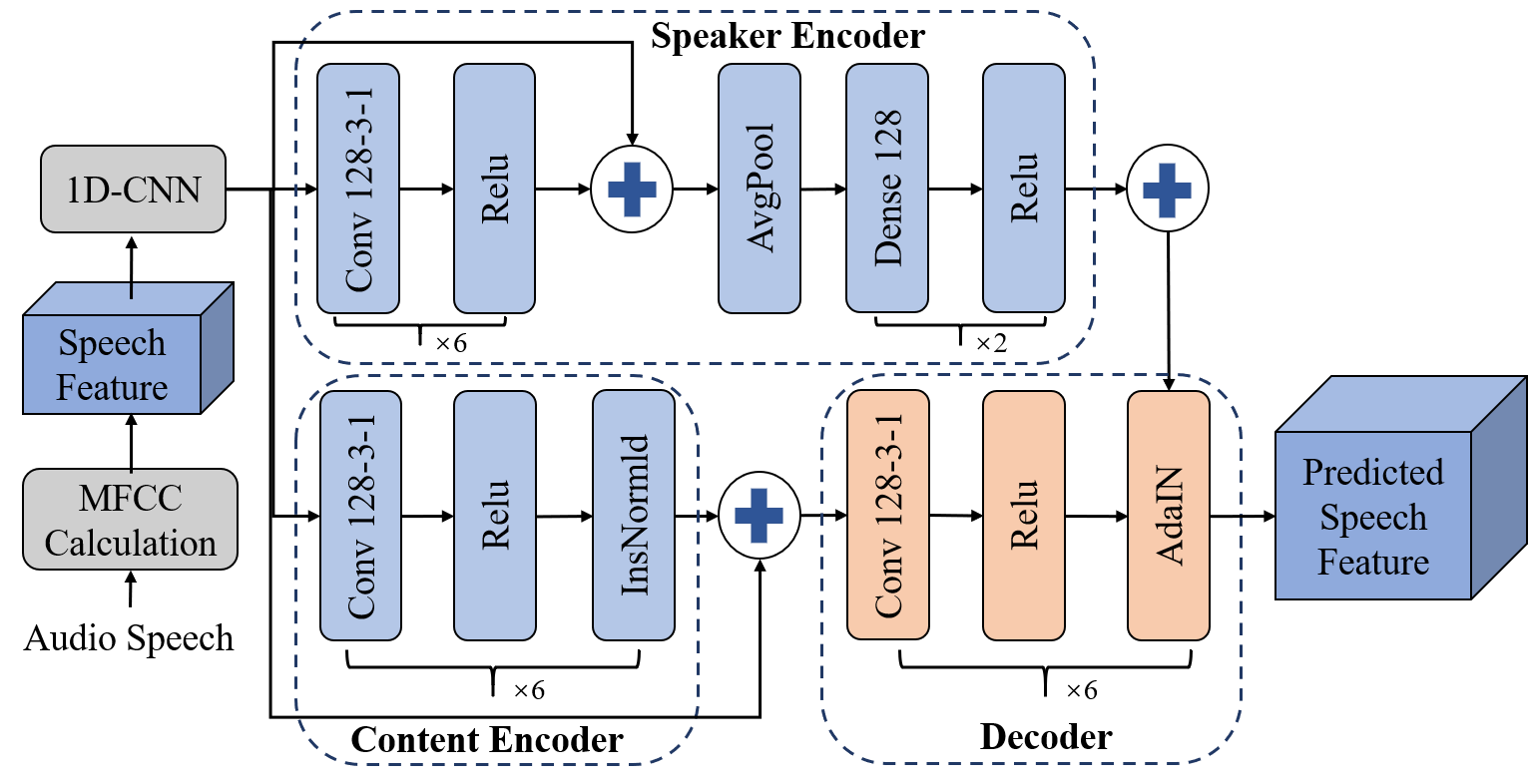}}

\end{minipage}
  	\caption{ Structure of the speech decomposition network.}
 	\label{fig:figure3}	
\end{figure}

\vspace{-0.2cm}
It was shown in \cite{C16} that speech signals inherently carry both non-linguistic information and linguistic information. The non-linguistic part refer to speaker identity, which is time-independent. While the linguistic part refer to content, which changes dramatically every several frames. On this basis, we pre-train the SDN (see Fig. \ref{fig:figure3}) to obtain the representation of speaker and content. Then the two representations obtained by the SDN, are fed as prior knowledge to adapt to the multi-speaker case, further improving the speaker generalization ability of SAFN. 

The SDN is a self-supervised model, composed of a speaker encoder, a content encoder and a decoder. The speaker encoder is trained to encode the non-linguistic information into the speaker representation. The content encoder is trained to encode the linguistic information into the content representation. And then the decoder is aimed to synthesize the speech feature by combining these two representations. However, we are just concerned about the part of speaker representation and content representation, so we pretrain the SDN  beforehand to obtain the speaker encoder and content encoder. Moreover, the SDN is trained on the whole dataset without using the articulatory labels. Thus, the proposed method is speaker-independent. We consider that the SDN has learned all the speakers acoustic information including identify information and content information. 

% \begin{figure}[h]
% 	\centering
% 	\includegraphics[width=0.95\linewidth]{ff2}
% 	\caption{ The module structure of the speech decomposition network.}
% 	\label{fig:figure3}	
% \end{figure}

The core of SDN is that, by normalizing the channel statistics which control the global information, the instance normalization (IN) \cite{C32} enforces the content encoder to focus on the linguistic part and remove the global information (\textit{i.e}, speaker information), while the averge-pool enforces the speaker encoder to focus on the non-linguistic part and learn the global information. Besides, the convolutional layer is used to capture long-term information. The dense layer \cite{C37} is used to enhance feature reuse and network training. And the adaptive instance normalization (AdaIN) \cite{C33} is utilized in decoder to bring the global information to the predicted speech feature though the corresponding parameters provided by speaker encoder. By doing this, the global information needed in the decoder is controlled by the speaker encoder. Thus, the SDN is encouraged to learn factorized representations. The loss of SDN is the reconstruction L1 loss between the input speech feature and the predicted speech feature. IN is expressed as:
\begin{equation}
\setlength{\abovedisplayskip}{3pt}
\setlength{\belowdisplayskip}{3pt}
\label{eq1}
M_{c}^{'}=\frac{M_{c}[w]-u_{c}}{\sigma_{c}},
\end{equation}
%\end{scriptsize}
where $ M_{c}$ represents the $c$-th channel with dimension $w$. $M_{c}[w]$ is the $w$-th element in $M_{c}$. To obtain IN, we first compute the mean $u_{c}=\frac{1}{W}\sum_{w=1}^{W}M_{c}[w]$, the standard variation $\sigma_{c}=\sqrt{\frac{1}{W}\sum_{w=1}^{W}(M_{c}[w]-u_{c})^2+\epsilon}$, where $\epsilon$ is a small value to avoid numerical instability. Each element in the array $M_{c}$ is normalized into $M_{c}^{'}$. 

% \begin{equation}\label{eq2}
% u_{c}=\frac{1}{W}\sum_{w=1}^{W}M_{c}[w] ,
% \end{equation}
%\end{scriptsize}

%\begin{scriptsize}
% \begin{equation}\label{eq3}
% \sigma_{c}=\sqrt{\frac{1}{W}\sum_{w=1}^{W}(M_{c}[w]-u_{c})^2+\epsilon},
% \end{equation}
%\end{scriptsize}
\vspace{-0.3cm}
\subsection{Auxiliary Feature Network}
\vspace{-0.1cm}
A new auxiliary feature is defined based on the EMA dataset \cite{C22}, where parallel acoustic and articulatory data are collected. Multiple sensors are attached to the pre-specified positions in EMA recordings. There are totally six sensors placed on the six articulators, namely tongue tip (T1), tongue blade (T2), tongue rear (T3), upper lip (UL), lower lip (LL), and lower incisors (LI). In particular, we divide these positions into two categories: outside visible positions (UL, LL, and LI) called lip auxiliary features in this work, and inner invisible positions (T1, T2, and T3) called the movements of tongue organ.

Then the above lip auxiliary features are estimated from the audio speech by AFN, instead of using the true label directly detected by the sensors. The reason is that we want to keep the same input (only acoustic speech) as the previous works. Besides, we do not freeze the parameters of the AFN during the training of the whole network.

The AFN is to estimate the lip auxiliary features from the personalized speech features, which is composed of three BLSTM layers to extract the contextual information, and two FC layers are followed by the BLSTM layers to generate trajectories of lip auxiliary features. The core of the AFN is expressed as:
\begin{equation}
\setlength{\abovedisplayskip}{3pt}
O_{t}^{l}= \sigma ( W_{io}^{l}x_{t} + b_{io}^{l} + W_{ho}^{l}h_{t-1} + b_{ho}^{l}), \notag
\end{equation}
\begin{equation}
O_{t}^{r}= \sigma ( W_{io}^{r}x_{t} + b_{io}^{r} + W_{ho}^{r}h_{t+1} + b_{ho}^{r}), \notag
\end{equation}
\begin{equation}
O_{t}=\dfrac{1}{2}\times(O_{t}^{l}+O_{t}^{r}),
\end{equation}
where $O_{t}^{l}$ is the lip auxiliary features estimated at the frame $t$ and the frame $t-1$, $O_{t}^{r}$ is the lip auxiliary features estimated at the frame $t$ and the frame $t+1$. ${x}_{t}$ is the input personalized speech features at frame $t$, $h_{t}$ is the temporary state at frame $t$ and $W_{io}^{l},b_{io}^{l}$ are the corresponding transformation matrix and bias from $i$ to $o$. By processing the forward and backward iteration, we obtain the $O_{t}$, which is the lip auxiliary features estimated at contextual information.

\vspace{-0.3cm}
\section{EXPERIMENTS}
\vspace{-0.35cm}
\subsection{Experimental Setup}
\vspace{-0.1cm}
\textbf{Datasets} The public MOCHA-TIMIT \cite{C35}, MNGU0 \cite{C21}, and HPRC \cite{C36} speech corpora include six reading locations set on T1, T2, T3, UL, LL and LI. In this work, we use the three tongue locations of X and Z directions (\textit{i.e}., T1, T2, T3) as our experimental predicted target. The MOCHA-TIMIT dataset consists of 460 utterances and EMA data recorded for one male and one female speaker, who speak British English. The MNGU0 dataset consists of 1263 utterances and EMA data recorded for one male speaking British English. The HPRC dataset consists of 720 utterances and EMA data recorded for eight native American English speakers.

\textbf{Performance Metrics} The performance is evaluated by two classical metrics, \textit{i.e}., root mean square error (RMSE) and correlation coefficient (CC) \cite{C11}. 

% The first criterion reports the mean deviation between estimated and the ground-truth trajectories, and the latter measures the similarity of the two trajectories. The measures are:
% \begin{equation}
% RMSE = \sqrt{\frac{1}{N}\sum_{i=1}^{N}(y_{i}-f(x_{i}))^{2}},
% \end{equation}
% \begin{equation}
% CC = \frac{\sum_{i=1}^{N}(f(x_{i})-\overline{f(x)})(y_{i}-\bar{y})}
% {\sqrt{\sum_{i=1}^{N}(f(x_{i})-\overline{f(x)})^{2}\sum_{i=1}^{N}(y_{i}-\bar{y})^{2}}},
% \end{equation}
% where $y_{i}$ and $f(x_{i})$ are the ground-truth and estimated EMA
% values of the $i^{th}$ frame, respectively; $\bar{y}$ and $\overline{f(x)}$ are mean values of $y_{i}$ and $f(x_{i})$.

\textbf{Implementation Details} In addition to the modules described in section \ref{sec:method}, the AIN contains three BLSTM layers with 100 units in each layer, followed by 2 FC layers. We train the proposed SAFN for 28800 iterations by Adam optimizer with a 1e-4 learning rate and the batch size is set as 5. Besides, the SOTA in \cite{C30} is reproduced as the baseline of our experiments. The SDN (shown in Fig. \ref{fig:figure3}) is pre-trained beforehand by Adam optimizer with a 5e-4 learning rate and the batch size is set as 25. Datasets are divided into the training set, the validation set, and the test set according to the proportion 8:1:1, respectively.	
\vspace{-0.3cm}
\subsection{Comparisons with the SOTA}
\vspace{-0.1cm}
To verify the generalization ability of SAFN, we conduct experiments by comparing the proposed SAFN with the SOTA \cite{C30} in two directions (\textit{i.e}., single speaker and multiple speakers) and four scenarios according to Table \ref{tab:TAB1}. S1 represents the experiment on single speaker. S2 represents the experiment on multi-speakers. S3 represents the speaker adaption experiment. More precisely, we first pool the training data from the whole dataset except the target speaker data to train a generic model. Then we fine tune the generic model weights using the target speaker data. S4 represents the speaker-independent experiment. The RMSE and CC in four scenarios are shown in Table \ref{tab:TAB3}.
% S1 means single speaker, S2 means multi-speaker, S3 means speaker adaptation and S4 means speaker-independent.
\vspace{-0.5cm}
\begin{table}[h]
	\caption{ Experimental setup for 4 different scenarios. S1 means single speaker, S2 means multi-speaker, S3 means speaker adaptation and S4 means speaker-independent. * means taking the corresponding proportion of data from each speaker. G, M and H represent dataset MNGU0, MOCHA and HPRC, respectively. -{}-{}- means no action.}%标题
	\label{tab:TAB1}
	\centering 
	\begin{threeparttable}
		\scriptsize
		%\centering%把表居中
		%\begin{spacing}{0.8}
			\setlength{\tabcolsep}{1.8mm}{
				\begin{tabular}{ccccccc}
					\toprule
					\textbf{Scenarios}&\textbf{Dataset}&\textbf{\#Speaker}&\textbf{Train}&\textbf{Validation}&\textbf{Fine-tune}&\textbf{Test}\\
					\midrule
					\textbf{S1}& G, M, H&1&80\%&10\%&-{}-{}-&10\%\\ 
					\midrule
					\textbf{S2}&H&N&80\%*&10\%&-{}-{}-&10\%\\ 
					\midrule
					\multirow{2}{*}{\textbf{S3}}&\multirow{2}{*}{H}&N-1&80\%*&20\%*&-{}-{}-&-{}-{}-\\
					&&1&-{}-{}-&-{}-{}-&80\%&20\%\\
					\midrule
					\multirow{2}{*}{\textbf{S4}}&\multirow{2}{*}{H}&N-1&80\%*&20\%*&-{}-{}-&-{}-{}-\\
					&&1&-{}-{}-&-{}-{}-&-{}-{}-&100\%\\
					\bottomrule
			\end{tabular}}
		%\end{spacing}
	\end{threeparttable}
\end{table}

\vspace{-0.8cm}
\begin{table}[h]
\caption{RMSE and CC for SOTA and SAFN in four scenarios.}%标题
		\label{tab:TAB3}
	\centering
	\begin{threeparttable}
		\centering%把表居中
		\scriptsize
		\begin{spacing}{0.85}
			\setlength{\tabcolsep}{1.2mm}{
				\begin{tabular}{ccccccccccc}%四个c代表该表一共四列，内容全部居中
					\toprule%第一道横线
					%			\multicolumn{2}{c}{Architecture}&$t_{1x}$&$t_{1y}$&$t_{2x}$&$t_{2y}$&$t_{3x}$&$t_{3y}$&RMSE&CC \\
					\textbf{Scenarios}&\textbf{Model}&\textbf{$t_{1x}$}&\textbf{$t_{1z}$}&\textbf{$t_{2x}$}&\textbf{$t_{2z}$}&\textbf{$t_{3x}$}&\textbf{$t_{3z}$}&\textbf{RMSE}&\textbf{CC} \\
					\midrule%第二道横线 
					\multirow{2}{*}{\textbf{S1(G)}}&SOTA&0.886&0.792&1.061&0.707&1.106&0.911&1.014&0.922 \\
					&SAFN&0.789&0.738&0.990&0.619&1.051&0.796&\textbf{0.830}&\textbf{0.941}\\
					\midrule%第二道横线
					\multirow{2}{*}{\textbf{S1(M)}}&SOTA&1.520&1.868&1.869&1.568&1.535&1.822&1.697&0.906\\
					&SAFN&1.289&1.497&1.403&1.442&1.571&1.551&\textbf{1.459}&\textbf{0.924}\\
					\midrule%第二道横线
					\multirow{2}{*}{\textbf{S1(H)}}&SOTA&1.725&1.760&1.871&1.684&2.060&2.170&1.881&0.901\\
					&SAFN&1.419&1.530&1.601&1.552&1.559&1.443&\textbf{1.517}&\textbf{0.922}\\
					\midrule%第二道横线
					\multirow{2}{*}{\textbf{S2}}&SOTA&1.730&1.840&1.901&1.721&2.100&2.210&1.917&0.890\\
					&SAFN&1.488&1.857&1.701&1.631&1.709&1.589&\textbf{1.662}&\textbf{0.903}\\
					\midrule%第二道横线
					\multirow{2}{*}{\textbf{S3}}&SOTA&1.730&1.759&1.830&1.651&2.030&1.850&1.808&0.911\\
					&SAFN&1.411&1.509&1.551&1.563&1.534&1.535&\textbf{1.507}&\textbf{0.925}\\
					\midrule%第二道横线
					\multirow{2}{*}{\textbf{S4}}&SOTA&2.675&3.803&3.384&2.102&2.878&3.227&3.009&0.701\\
					&SAFN&2.184&3.077&2.938&2.621&2.412&3.096&\textbf{2.721}&\textbf{0.751}\\
					\bottomrule%第三道横线
			\end{tabular}}
			%		\begin{tablenotes}
			%
			%		\end{tablenotes}
		\end{spacing}
	\end{threeparttable}
\end{table}

%轨迹图片
% \begin{figure}[htb]
% \begin{minipage}[b]{1.0\linewidth}
%   \centering
%   \centerline{\includegraphics[width=8.5cm]{draw1}}
% %  \vspace{2.0cm}
% \end{minipage}
%   	\caption{Comparisons of the predicted results of the tongue movement between SAFN and SOTA. In each box, the blue line is real label, the red dashed line is predicted by SAFN and the green dashed line is predicted by SOTA method.}
%  	\label{fig:figure1}	
% \end{figure}

% \begin{figure}[h]
% 	\centering
% 	\includegraphics[width=0.9\linewidth]{draw1}
% 	\caption{Comparisons of the predicted results of the tongue movement between SAFN and SOTA. In each box, the blue line is real label, the red dashed line is predicted by SAFN and the green dashed line is predicted by SOTA method.}
% 	\label{fig:figure1}	
% \end{figure}
\vspace{-0.22cm}
Table \ref{tab:TAB3} shows RMSE and CC value in four scenarios among three public datasets. Basically, we can observe that the proposed SAFN outperforms SOTA by almost 0.18mm $\sim$ 0.36mm on RMSE. Besides, CC scores show a similar trend (increase by 1.4\% $\sim$ 5.0\%). It is obvious to see that the improvement of CC in the speaker-independent case (S4) is 5.0\%, which is much higher than that in the speaker-dependent cases (S1, S2 and S3). It indicates that the prior speaker identity features obtained by SDN can effectively alleviate the mismatch between the acoustic space of the speakers in the training set and those in the test set, further to adapt to the speaker-independent case. 

%  Qualiative comparisons between SAFN and SOTA are shown in Fig. \ref{fig:figure1}. It can be clearly observed that the predicted tongue articulators generated by SAFN are more accurate than SOTA, especially in the red circle of the Fig. \ref{fig:figure1}. It indicates that the prior information (\textit{i.e}., lip auxiliary information, content information and speaker information) have a positive effect on the reconstruction of tongue articulatory movements.
\vspace{-0.3cm}
\subsection{Ablation Study}
\begin{figure}[htb]
\begin{minipage}[b]{1.0\linewidth}
  \centering
  \centerline{\includegraphics[width=8cm,height=5cm]{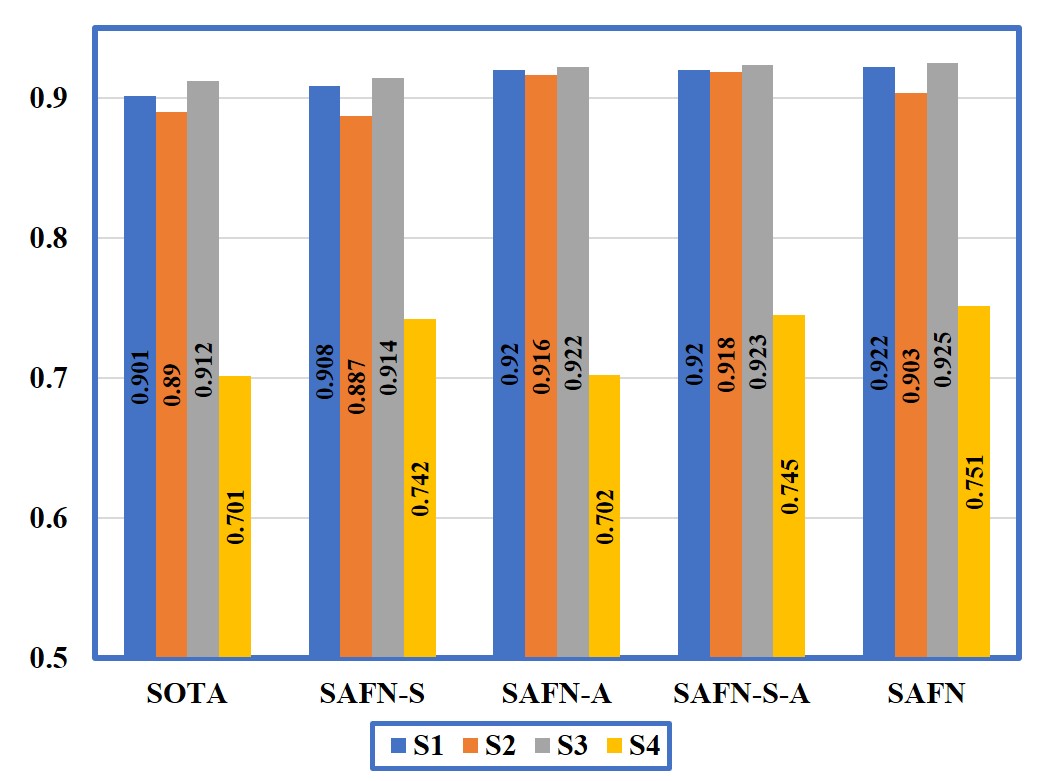}}
\end{minipage}
  	\caption{CC of the different neural network in four scenarios.}
 	\label{fig:ff4}	
\end{figure}

\vspace{-0.1cm}
To verify the effectiveness of the proposed modules in Section \ref{sec:method}, we carry out the ablation experiment according to Table \ref{tab:TAB2}, and the results are shown in Fig. \ref{fig:ff4}.
\vspace{-0.3cm}
\begin{table}[h]
	\scriptsize
	\caption{\textbf\tiny{Ablation experiment verifies the performance of each module. SAFN-S, SAFN-A and SAFN-S-A represents the AIN comparing with SDN, AFN and both of the above two parts, respectively.}}%标题
	\centering
	\label{tab:TAB2} 
	\begin{threeparttable}
		\begin{spacing}{0.8}
			\setlength{\tabcolsep}{7.2mm}{
				\begin{tabular}{cccc}
					\toprule
					\textbf{}&\textbf{SDN}&\textbf{AFN}&\textbf{FTN}\\
					\midrule
					\textbf{SOTA}&\textbf{$\times$}&\textbf{$\times$}&\textbf{$\times$}\\
					%\midrule
					\textbf{SAFN-S}&\textbf{\checkmark}&\textbf{$\times$}&\textbf{$\times$}\\
					%\midrule
					\textbf{SAFN-A}&\textbf{$\times$}&\textbf{\checkmark}&\textbf{$\times$}\\
					%\midrule
					\textbf{SAFN-S-A}&\textbf{\checkmark}&\textbf{\checkmark}&\textbf{$\times$}\\
					%\midrule
					\textbf{SAFN}&\textbf{\checkmark}&\textbf{\checkmark}&\textbf{\checkmark}\\
					\bottomrule
			\end{tabular}}
		\end{spacing}
	\end{threeparttable}
\end{table}
\vspace{-0.6cm}
% \begin{figure}[h]
% 	\centering
% 	\includegraphics[width=0.8\linewidth]{figure4}
% 	\caption{CC of the different neural network in four scenarios.}
% 	\label{fig:ff4}	
% \end{figure}
\vspace{0.8cm}
Obviously, in the speaker-dependent cases (S1, S2 and S3), those models with AFN (\textit{i.e}., SAFN, SAFN-S-A and SAFN-A) outperform those without AFN (\textit{i.e}., SOTA and SAFN-S). However, in the speaker-independent case (S4), those models with SDN (\textit{i.e}., SAFN, SAFN-S-A and SAFN-S) outperform those without SDN (\textit{i.e}., SOTA, SAFN-A). Based on the above results, it is demonstrated that in the speaker-dependent case, the AFN improves the performance by adding lip auxiliary features as prior knowledge. In the speaker-independent case, the prior personalized speaker information obtained by SDN brings a large gain. Besides, comparing SAFN with SAFN-S-A, we hypothesis that FTN improves the performance of SAFN by enhancing the correlations between the personalized speech features and the lip auxiliary features. 
% The paper title (on the first page) should begin 1.38 inches (35 mm) from the
% top edge of the page, centered, completely capitalized, and in Times 14-point,
% boldface type.  The authors' name(s) and affiliation(s) appear below the title
% in capital and lower case letters.  Papers with multiple authors and
% affiliations may require two or more lines for this information. Please note
% that papers should not be submitted blind; include the authors' names on the
% PDF.
\vspace{-0.730cm}
\section{Conclusion}
\vspace{-0.35cm}
In this work, we propose a novel network SAFN to promote the generalization ability of speaker-independent AAI and further improve AAI performance. Firstly, to improve the generalization ability of the proposed SAFN, a SDN is presented to obtain the speaker embedding and content embedding as the personalized speech features. Besides, to further improve the performance of AAI, a new AFN is proposed to obtain the lip auxiliary features as prior knowledge to help the prediction of the tongue organ. Experimental results on three public datasets demonstrate that both in speaker-dependent and speaker-independent scenarios, the SAFN outperforms SOTA by a large margin. For the future work, the self-supervised method based on Meta Learning will be applied to the speaker-independent AAI task.

\vspace{-0.7cm}
\section{ACKNOWLEDGEMENTS}
\vspace{-0.25cm}
This work was supported by the National Natural Science Foundation of China(No. 61977049), National Natural Science Foundation of China (No. 62101351) and the Tianjin Key Laboratory of Advanced Networking.
\bibliographystyle{IEEEbib}
\bibliography{mybib}
\end{document}